 \documentclass{aastex}

\shortauthors{D.~K. Galloway}
\shorttitle{Comptonization in the X-ray pulsar GX~1+4}

\begin{document}




\title{Comptonization in the accretion column of the X-ray pulsar GX~1+4}


\author{D.~K. Galloway\altaffilmark{1}}
\affil{School of Mathematics and Physics, University of Tasmania, Hobart,
Australia 7001}
\affil{RCfTA, University of Sydney, Camperdown, NSW 2006}


\altaffiltext{1}{present address: Center for Space Research, MIT,
    77 Massachusetts Avenue, Cambridge, MA 02139-4307}


\begin{abstract}

X-ray observations of the binary pulsar GX~1+4 made using the Rossi X-ray
Timing Explorer ({\it RXTE\/}) satellite between February 1996 and May 1997
were analysed to quantify source spectral variation with luminosity.

Mean Proportional Counter Array (PCA) spectra over the range
2--40~keV are best fitted with a Comptonization model, with source spectrum
temperature $T_0\approx 1$--1.3~keV, plasma temperature $T_{\rm
e}\approx 6$--10~keV, and optical depth $\tau\approx 2$--6.  The range of
fitted $T_0$ was consistent with the source spectrum originating at the
neutron star polar
cap, with Compton scattering taking place primarily in the hot plasma of the
accretion column.  Both the fitted optical depth and plasma temperature vary
significantly with the source flux. The variation of the optical depth
(and hence the density of the scattering region) with luminosity
strongly suggests increasing cross-sectional area of the
accretion column at higher accretion rate $\dot{M}$.

The wide range of source luminosity
spanned by archival observations of GX~1+4 offers evidence for two distinct
spectral states above and below $L_{\rm X}\approx2\times10^{37}\ {\rm
erg\,s}^{-1}$ (2--60~keV, assuming a source distance of 10~kpc).  GX 1+4
additionally exhibits dramatic hourly variations in neutral column density
$n_{\rm H}$ indicative of density variations in the stellar wind from the
giant companion.

\end{abstract}




\section{Introduction}

At the time of its discovery \cite[]{lew71} GX~1+4 was one of the brightest
objects in the X-ray sky.  The optical counterpart is an M6 giant
\cite[]{dav77} which is readily classified as a symbiotic binary
\cite[]{symcat00}.  Unusually this LMXB appears to be accreting from the
companions' stellar wind, which infrared observations suggest is at least an
order of magnitude faster than typical red giant winds \cite[]{chak98:ir}
and which presumably also gives rise to the optical emission line nebula.
Optical spectroscopy suggests a source distance of 3--15~kpc
\cite[]{chak97:opt}.

Measurements of the average spin-up rate during the 1970s gave the largest
value recorded for any pulsar (or in fact any astronomical object) at
$\approx 2$ per cent per year. Inexplicably, the average spin-up trend
reversed around 1983, switching to spin-down at approximately the same rate.
Since that reversal a number of changes in the sign of the torque (as
inferred from the rate of change of the pulsar spin period) have been
observed \cite[]{chak97}.  In contrast to predictions of the
\cite{gl79a,gl79b} model BATSE measurements suggest that the torque measured
for GX~1+4 is sometimes {\it anticorrelated} with luminosity
\cite[]{chak97}.  This behaviour has not been observed in other pulsars.
The orbital period is not known, but secular torque variations suggest
$P_{\rm orb}\approx304$~d \cite[]{cut86,per99}.  Several estimates
\cite[]{beu84,dot89,mony91,gre93,cui97} indicate a neutron star surface
magnetic field strength of 2--3$\times 10^{13}$~G.

The X-ray flux from the source is extremely variable on time-scales of
seconds to decades. Two principal flux states have been suggested, a `high'
state which persisted during the spin-up interval of the 1970s, and a `low'
state since.
Superimposed on the long-term flux variations are smooth changes on
time-scales of order hours to days. On the shortest time-scales the periodic
variation due to the neutron star's rotation period at around 2 min is
observed.  Pulse profiles from the source exhibit a bewildering variety of
shapes.  The profiles are typically asymmetric, with a
sharp dip forming the primary minimum
\cite[][, Galloway \& Greenhill 2000, in preparation]{dot89,gre98,kot99}.
These dips appear to be signatures of near-field `eclipses' of the polar cap by
the accretion stream (Galloway et al. 2000, in preparation).
The sense of asymmetry of the
profiles is variable on timescales as short as a few hours \cite[]{gil00}.

Historically the X-ray spectrum has been fitted with bremsstrahlung or power
law models, however more recent observations with improved spectral resolution
generally favour a power law model with exponential cutoff.
In general the various power-law
models are an approximation to a spectrum originating from a
low-energy distribution of photons Comptonized by scattering within a hotter
plasma \cite[e.g.][]{pss83}. The accretion column plasma reaches the highest
density immediately above the polar cap, where the X-ray photons are thought
to be emitted. Compton scattering in the accretion column
is likely to be an important process affecting the initial photon spectrum.
More detailed approximations to Comptonized spectra are an obvious
choice for candidate X-ray pulsar models; the \cite{tit94} formulation
has previously been used in fits to {\it RXTE} spectra \cite[]{gal00:spec}.
For any spectral model covering the range 1-10~keV, it is also
necessary to include a gaussian component representing iron line emission at
$\approx 6.4$~keV, and a term to account for the effects of photoelectric
absorption by cold gas along the line of sight with hydrogen column density
in the range $n_H = (4-140) \times 10^{22}\ {\rm cm^{-2}}$.
The source spectrum and in particular the column density $n_H$ have previously
exhibited significant variability on time-scales of a day or less
\cite[]{beck76,gal00:spec}.

The public archival data now available from {\it RXTE} observations of the
source has allowed a test of the applicability of Comptonization
models to spectra over a wide range of source luminosities.

\section{Observations}

{\em RXTE} consists of three instruments, the proportional counter array
(PCA) sensitive to photons in the energy range 2--60~keV, the high-energy
X-ray timing experiment (HEXTE) covering 16--250~keV, and the all-sky
monitor (ASM) which spans 2--10~keV \cite[]{giles95}.
Pointed observations are performed using
the PCA and HEXTE instruments, while the ASM regularly scans the entire
visible sky. During targeted observations, interruptions may be made for
previously scheduled monitoring or target-of-opportunity (TOO) observations
of other sources.

Six of the eight event analysers (EAs) aboard {\it RXTE} are available for
processing of events measured by the PCA. Two EAs are dedicated to modes
which are always present, `Standard-1' and `Standard-2'.  The Standard-1
mode features $0.25\ \mu{\rm s}$ time resolution but only one spectral
channel, while Standard-2 offers 129 spectral channels between 0--100~keV
(the sensitivity above 60~keV is negligible) with binning on 16~s time
resolution. The remaining four EAs may be used in a range of
user-configurable modes, which are in general only necessary to obtain
spectra on a finer time resolution.

\begin{figure}
\plotone{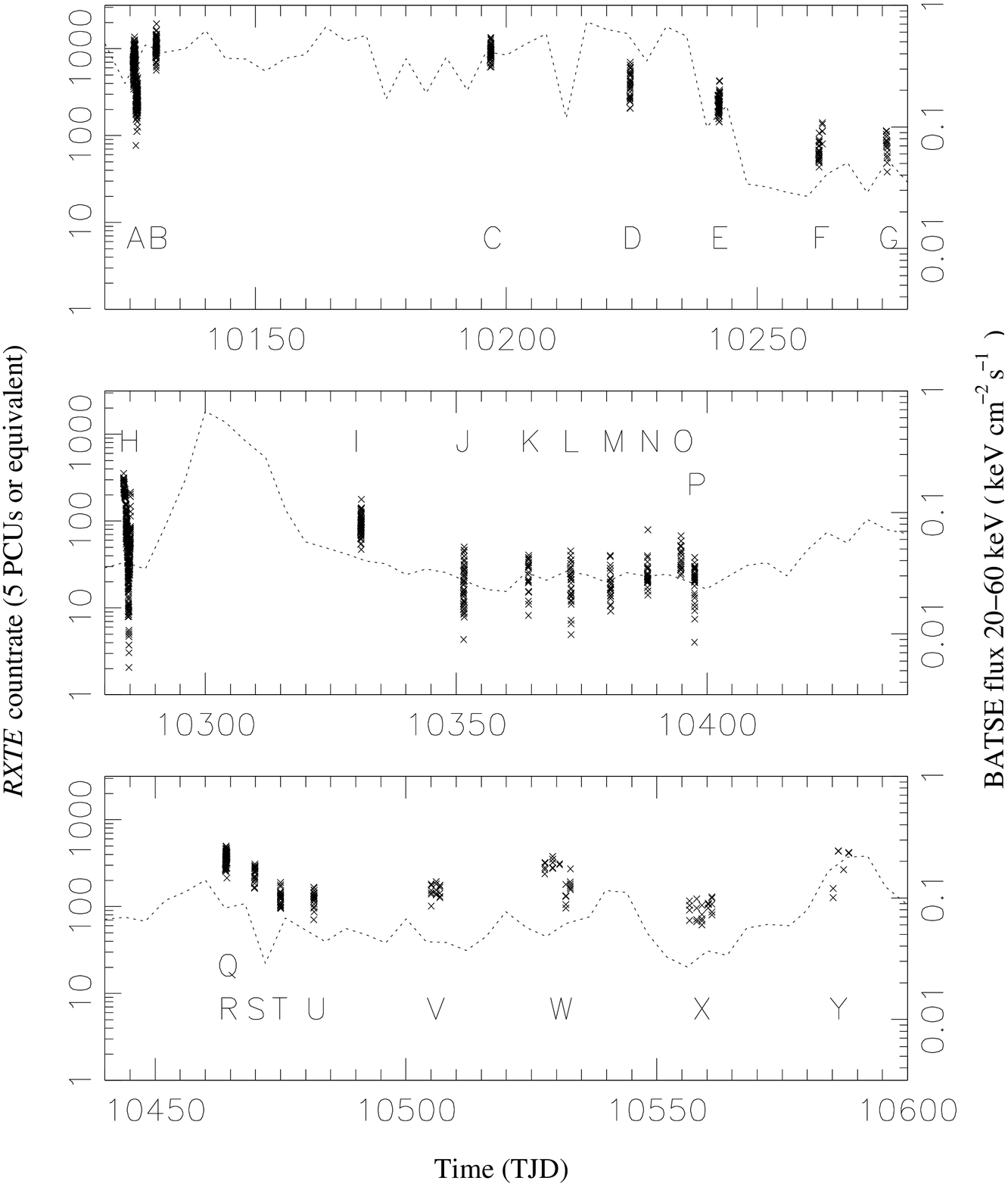}
\caption[flux.eps]{ {\it RXTE} and BATSE observations of GX~1+4
between Feburary 1996
and May 1997. Phase-averaged background-subtracted countrate from pointed
observations using the PCA aboard {\it RXTE} are plotted as crosses (left hand
scale in each panel).  The observations are labelled A$\ldots$Y; see Table
\ref{tab-gxobs}.  The dotted line represents the 20--60~keV pulsed flux as
measured by BATSE (right hand scale).
\label{gxrate}}
\end{figure}

Between February 1996 and May 1997 GX~1+4 was observed by {\it RXTE} in 25
distinct intervals totalling $\approx 230$~ks (Table \ref{tab-gxobs}).
%
%
From mid-1996 onwards two of the five proportional counter units (PCUs) which
make up the PCA were
intermittently turned off to partially arrest a gradual sensitivity
degradation measured by the {\it RXTE} team.  All 5 PCUs were on for the
majority of the observations. PCUs 3 and 4 were commanded off during
observations D, N and P, and for part of observation H.  PCU 4 alone was off
for short intervals during observations H and R.  The countrate for these
observations was scaled appropriately to give the equivalent for all 5 PCUs on.
The phase-averaged countrate from the source demonstrates its dramatic
variability (Figure \ref{gxrate}).  
The pulse period for the averaging has been determined by
phase folding or interpolation of BATSE period measurements; see 
Galloway \& Greenhill 2000, in preparation.  The countrate can
span more than two orders of magnitude in less than two days
\cite[e.g. observation H;][]{gil00}.  More typically the countrate varies
by a factor of $\sim 5$
on hourly timescales when the countrate was 100~ct$\ {\rm s}^{-1}$ or more,
and a factor of $\sim 10$ otherwise.  Increased variability at low flux
levels was also noted by \cite{gre99} in analyses of archival hard X-ray
observations.  These variations were superimposed on long-term trends with
timescales of a year or more, which may be related to the orbital period
\cite[]{per99}.  The source appeared to be relatively bright during the
first half of 1996, after which it entered a $\approx 50$~d low state. The
decrease in 20--60~keV flux was much more abrupt than the PCA countrate.
At this time pulsations from the source all but ceased, suggesting cessation
of mass transfer from the companion or possibly centrifugal inhibition of
accretion \cite[]{cui97}.  A flare lasting about 30 days was observed by
BATSE during August 1996, following the period of lowest flux during July.
After the flare the source returned to a relatively low state, which
persisted to June 1997.  The PCA countrate generally traces the 20--60~keV
flux as measured by BATSE reasonably well, but spectral variation in the
source (in particular the degree of low-energy absorption) means that the
contribution from high-energy photons varies.

\section{Spectral analysis}

\begin{figure}
\plotone{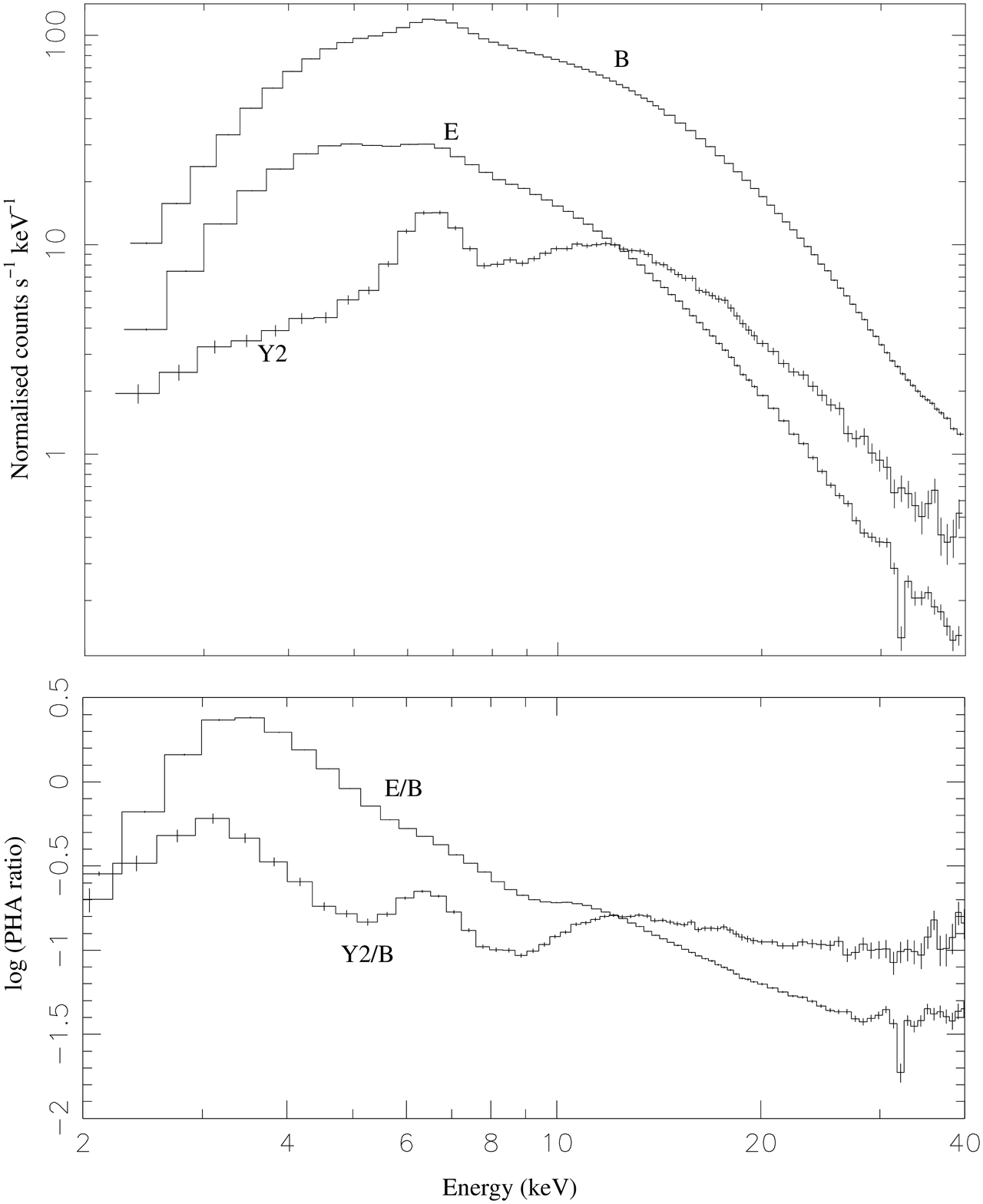}
\caption[spec-comp.eps]{ Sample raw {\it RXTE} spectra from GX~1+4.
The top panel plots mean
spectra from observations B (17 February 1996), E (8 June 1996) and Y2 (17 May
1997). The bottom panel shows the PHA ratio calculated by dividing the
countrate in each spectral bin for spectra from observations E and Y2 by the
spectrum from observation B.
\label{spec-comp}}
\end{figure}

Analysis of Standard-2 {\it RXTE} data presented in this paper was carried out
using {\sc lheasoft~5.0}, released 23 February 2000 by the {\it RXTE} Guest
Observer Facility (GOF). The data were first screened to ensure that the
pointing offset was $<0.02\arcdeg$ and the source was $>10\arcdeg$ from the
sun. This introduces additional gaps to the data.  If individual
PCUs are switched off or on during the observation it was necessary to
extract spectra from each interval with a constant number of PCU's on and
analyse them separately.
Observations were subdivided when significant countrate variations are
seen within the observation (e.g.  observations A and H).
As the spectral fit
parameters (particularly the column density $n_{\rm H}$) tend to vary on
hourly timescales it may be necessary to further subdivide intervals to
obtain adequate fits. Subintervals were labeled with the appropriate
letter from Table \ref{tab-gxobs} and a number (e.g. A1, A2 $\ldots$).
Example spectra are shown in Figure \ref{spec-comp}.

Instrumental background from cosmic ray interactions and as a result of
satellite passages close to the SAA were estimated using the {\sc pcabackest}
software which is included in the {\sc lheasoft} package.  The background
models used were also from the 23 February 2000 release. Where the net source
countrate was $\la 60\,{\rm count\,s}^{-1}$ the `faint' source models have
been used, with the `bright' models used otherwise.

Candidate spectral models were tested by fitting to the count rate spectrum
to minimise the reduced-$\chi^2$ ($=\chi^{2}_\nu$) using the {\sc xspec}
spectral fitting package version 11 \cite[]{xspec}.
{\sc xspec}
allows spectral fitting given some assumed level of {\em systematic} error
in the instrumental calibration, which clearly will affect the fit
statistic. However no systematic error was assumed in the spectral fits
undertaken in this paper.
In general each model takes the form of at least one continuum component
and a multiplicative component to account for absorption by cold matter
along the line of sight. Continuum components tested for the GX~1+4
spectra include powerlaw, broken powerlaw, powerlaw and blackbody and
powerlaw with exponential cutoff, as well as the
\cite{tit94} Comptonization approximation implemented in {\sc xspec} as
`{\tt compTT}'.  Parameters for the latter component include
the input spectrum temperature $T_0$ (approximated as a Wien law), the
scattering plasma temperature $T_{\rm e}$ and normalisation $A_{\rm C}$ (in
units of photons$\,{\rm cm^{-2}\,s^{-1} \,keV^{-1}}$).  Additionally the
redshift can be specified, but this was fixed at 0.0 for the model fits
described in this paper.
The effective optical depth $\tau$ is calculated from the spectrum assuming
either a spherical or planar (disc) geometry. If the accretion column
{\em is} the primary region where Compton scattering takes place,
clearly neither of these assumptions are strictly appropriate.
For the {\tt compTT} model fits presented in this paper I adopt the `disc'
geometry switch, noting that the calculated $\tau$ will thus be only an
approximate measure of the true optical depth in the column. In general the
calculated $\tau$ for an assumed spherical geometry will be approximately
twice as large.

A gaussian component with energy $\approx 6.4$~keV simulates the (generally)
significant iron line emission.  Due to its proximity to the galactic plane,
an additional component which takes into account the so-called `galactic
ridge' emission must also be included in the spectral model.  This component is
modelled using a Raymond-Smith emission spectrum from hot, diffuse gas and a
powerlaw, both absorbed by a neutral column density of $n_{\rm H}=2\times
10^{21}\,{\rm cm}^{-2}$. Two additional power law
components serve as a correction to the diffuse X-ray background assumed by
{\sc pcabackest}.  The model was identical to that fitted to survey data from
this region \cite{val98} with normalisations and metal abundance as fitted to
spectra taken during slews to and from the source in July 1996
\cite[observation H in Table \ref{tab-gxobs};][]{gal00:spec}.
A final instrumental effect which must be
taken into account to obtain the lowest possible residuals in the model fit
is a consequence of the Xenon L-edge absorption feature. This feature is
modelled by a multiplicative edge model component with energy fixed
(`frozen') at 4.83~keV.

\section{Results}

For any spectral model, the fit statistic $\chi^2_{\nu}$ depends strongly on
the countrate during the observation. When the countrate (and hence the
source-to-background ratio) is low, errorbars on the spectral bins will be
large, and typically more than one (if not all) of the models will give a
statistically acceptable spectral fit. The comparison between the models for
those observations at the highest available countrates (e.g. A$\ldots$E,
Q$\ldots$U) provides a much more stringent test. For the powerlaw continuum
during these intervals $\chi^2_{\nu}$ varies from $\approx 2$--150,
typically $\approx 5$. A powerlaw modified by an exponential high-energy
cutoff provides some improvement, with typical values $\approx 1.5$--6, as
does a broken power law (with two spectral indexes above and below some
`break' energy) giving $\chi^2_{\nu} \approx 1.5$--4.  A combination
powerlaw and blackbody was acceptable for most of the observations with
typical values $\approx 1.3$, but can be as high as e.g.
$\chi^2_{\nu}\approx25$ for observation A1. Only the Comptonization
continuum gives $\chi^2_{\nu} \la 2.5$ for each of the spectra tested.

\begin{figure}
\plotone{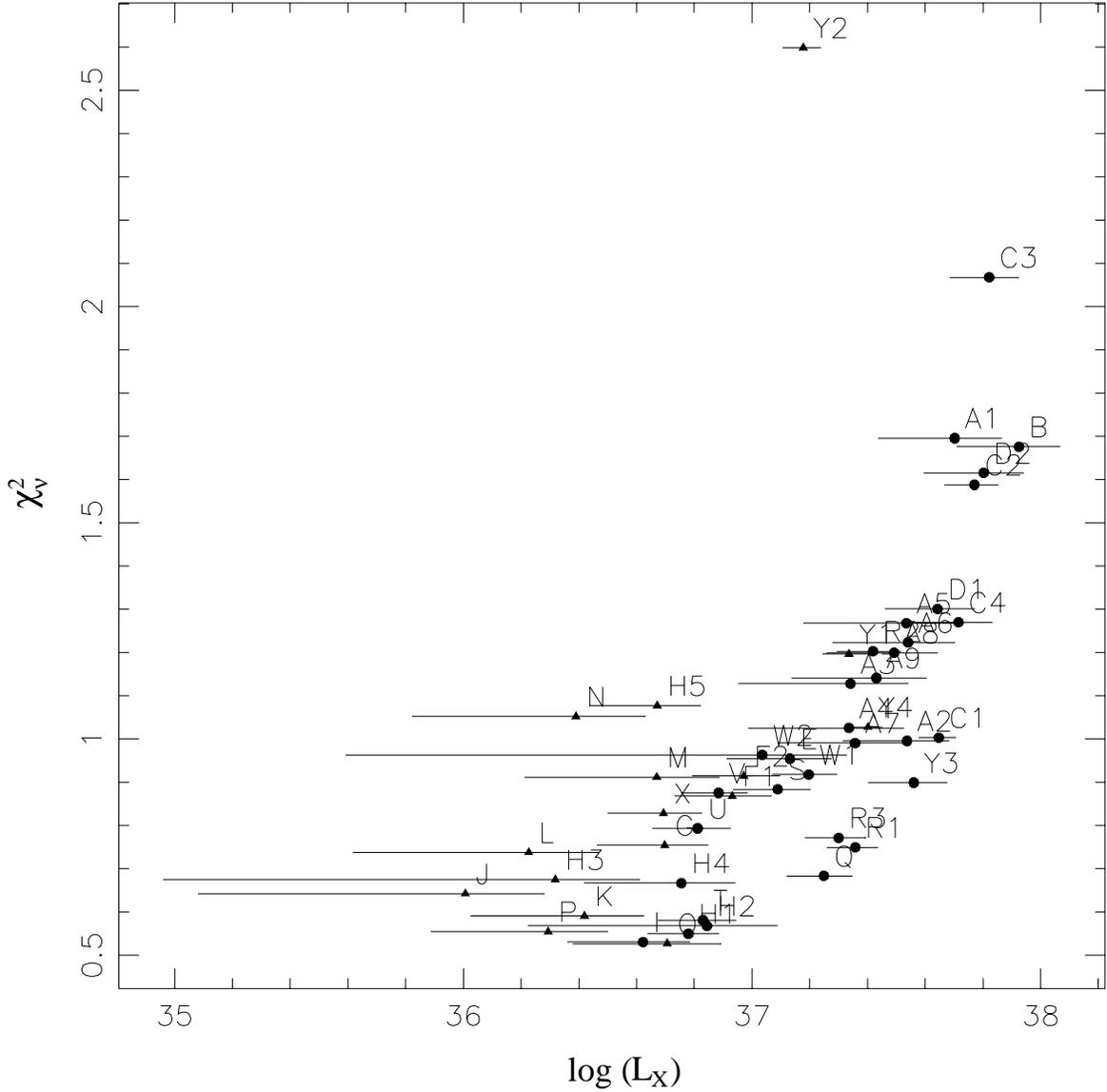}
\caption[chi-all.eps]{ Reduced-$\chi^2$ for fits to {\it RXTE} spectra from
GX~1+4 using
an analytic Comptonization continuum component ({\tt compTT} in {\sc xspec}) as
a function of 2--60~keV flux (assuming a source distance of 10~kpc). Spectral
range for fitting is typically 2.5--35~keV.  Circles indicate those fits where
all the spectral parameters are free to vary, while for observations marked by
triangles at least one parameter (usually $T_{\rm e}$) is fixed in order to
obtain error limits on the other parameters.  The markers are labelled with
the letter corresponding to the observation, in some cases also with a number
where the observation has been divided into separate intervals for spectral
analysis.  Errorbars show 90 per cent limits on flux calculated from the
variance in the phase-averaged countrate over each observation (see Figure
\ref{gxrate}). \label{gxchi}}
\end{figure}

The correlation between $\chi^2_{\nu}$ for the Comptonization component fits
and the integrated source flux is readily apparent
although there is also significant scatter about the line of fit (Figure
\ref{gxchi}). Significant variations in $\chi^2_{\nu}$ were seen even between
immediately adjacent observations (e.g. R1/R2/R3). While it was necessary to
subdivide certain intervals (most noticeably the second half of observation
A) to obtain reasonable values of $\chi^2_{\nu}$,
the Comptonization model still unambiguously offers the best fit over the
shorter integration times resulting. The spectrum extracted
from observation Y2 is a rather special case; see Galloway 2000, in
preparation.

The fitted values of the source spectrum temperature $T_0$ were seen to cluster
strongly around the weighted mean of $1.284 \pm 0.004$~keV (1$\sigma$)
over a wide range of flux (Figure \ref{gxcomp}a). Below $\approx 3\times
10^{36}\ {\rm erg\,s}^{-1}$ the fit values were subject to large
uncertainties but were broadly consistent with $T_0 \la 3$~keV. The fit
parameter ranges are limited at the lower boundary by the minimum allowed by
the {\tt compTT} implementation in {\sc xspec} of 0.001. Several spectra
suggest $T_0<1$~keV, however the 90 per cent error limits were large and these
may not be significant.

The fitted optical depth $\tau$ varies between 2--6 over the flux range of
$10^{36}$--$10^{38}\ {\rm erg\,s}^{-1}$ (Figure \ref{gxcomp}b). 
This figure omits the value for observation Y2, which was by far the
highest measured at around $\tau=19.4^{21.1}_{18.3}$.
When the flux was $\la 2\times 10^{37}\ {\rm erg \,s}^{-1}$ the $\tau$ values
cluster around 3, whereas above that a range of 4-6 is more typical.  The
corresponding mean values were $2.95\pm0.47$, and $4.7\pm1.0$ respectively,
which disagree at the 1.6$\sigma$ level. The variance was substantially
greater at high flux.  The observed variation may be viewed as a rough
correlation between $\tau$ and $L_{\rm X}$. To test the significance of the
relationship I use the BCES(Y|X) estimator of \cite{akritas96}. Ordinary
weighted least squares methods are not strictly appropriate for fitting data
with heteroscedastic (varying) errors in both parameters, and where there
are thought to be {\it systematic} errors in addition to measurement errors
in the parameters. The resulting gradient for a linear fit confirms a
significant relationship, with gradient $b=0.450\pm0.070$.  However,
separate linear fits to the data above and below $L_{\rm X}=2\times10^{37}\
{\rm erg\,s}^{-1}$ tell a different story, with gradients
$b=-0.0354\pm0.638, -1.41\pm1.01$ respectively.

It is possible that the variation of fitted $\tau$ values is due to aliasing
by some dependence of the fitting algorithm on the varying countrate. This
can be tested by (for example) selecting a spectral model with parameters
from one of the higher flux observations, generating a simulated spectra
with much lower flux, and comparing the resulting spectral fitting results.
In general the spectral parameters were found to depend only weakly on flux
for simulated observations. There was certainly no abrupt drop to lower
$\tau$ values as the simulated flux was decreased.  Visual inspection of
representative spectra from the high- and low-$\tau$ groups further supports
the significant differences between the two (Figure \ref{spec-comp}). Fits
to mean spectra from observations B and E give consistent column densities
$n_{\rm H}=5.51_{5.36}^{5.66}$ and $5.37_{5.00}^{5.72}$ respectively.
However, the peak in the spectra clearly occurs at different energies. The
PHA ratio between the two demonstrates that the high-energy slopes are
different, as is the turnover energy (where the slope changes around
10~keV; note the `kink' in the PHA ratio around that energy).

The corresponding plot for $T_{\rm e}$ also shows evidence for a
non-isotropic distribution of the points, but to a lesser extent (Figure
\ref{gxcomp}c).  The points appear to fall into two groups again roughly
divided at a flux level of $\approx 2\times 10^{37}\ {\rm erg \,s}^{-1}$.
Above this level $T_{\rm e}\la 10$~keV typically, whereas below $T_{\rm e}$
values were generally higher although with very large errorbars.  The
distribution may be viewed as an anticorrelation of $T_{\rm e}$ with flux;
the resulting BCES(Y|X) slope was $-1.76\pm0.54$.  Within the two groups
$T_{\rm e}$ appears to increase with flux, although no significant linear
relationship was found.  With low countrates it was much more difficult to
determine $T_{\rm e}$ precisely, and in fact for approximately half the
spectra which fall into this lower flux range the value cannot be determined
at all (note the smaller number of points in Figure \ref{gxcomp}c compared to
a,b). For these spectra I fix the value at 10~keV, consistent with the
mean for the group $10.6\pm2.1$. This was different from the mean
for the higher flux group, $7.8\pm1.2$, at low signifcance only ($\approx
1.2\sigma$).

The Compton $y$-parameter can be calculated for each spectra from the fitted
$\tau$, $T_{\rm e}$. Since $kT_{\rm e} \ll m_{\rm e}c^2$ the nonrelativistic
form is more appropriate \cite[]{rl79}.  The plot of $y$ as a function
of source flux exhibits similar clustering to the corresponding plots with
$\tau$, $T_{\rm e}$ (Figure \ref{gxcomp}d). The two groups fall neatly on
either side of the line $y=1$, normally considered the limiting value between
spectra which are weakly modified ($y\ll 1$) and spectra strongly modified
($y\gg 1$) by Comptonization.

\begin{figure}
\plotone{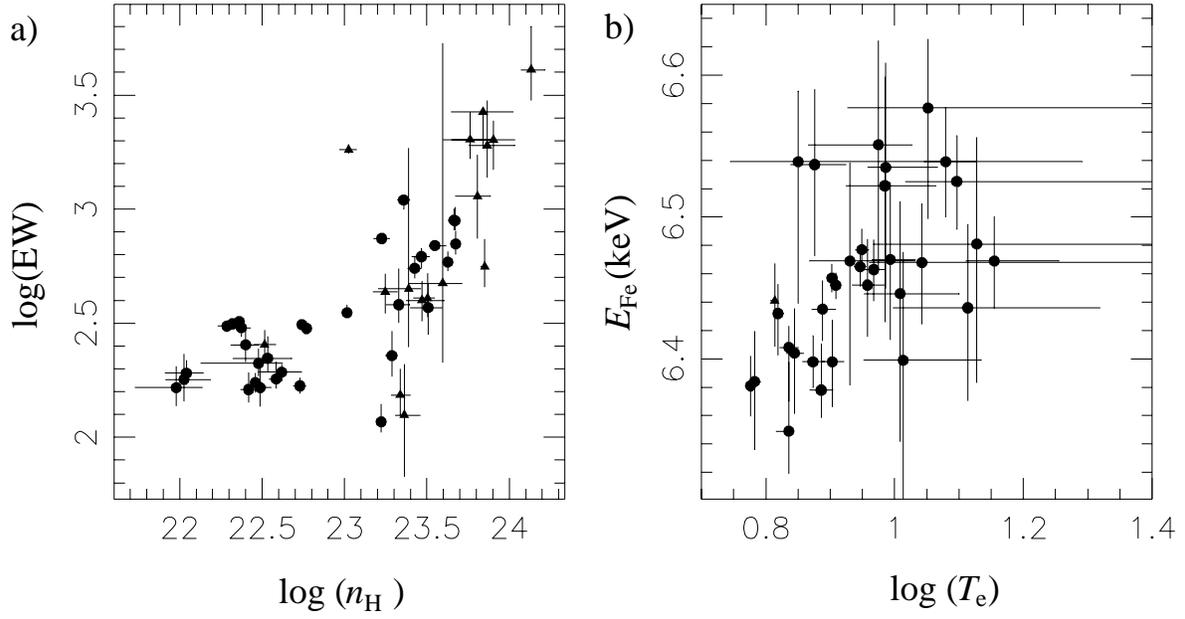}
\caption[line.eps]{a) Iron line equivalent width (EW) plotted
as a function of
absorption column density $n_{\rm H}$. b) Iron line gaussian centre energy
plotted as a function of the plasma temperature $T_{\rm e}$.
Errorbars show 90 per cent confidence limits. Other details are as for Figure
\ref{gxchi}.
\label{gxline} }
\end{figure}

The absorption column density $n_{\rm H}$ varies by more than two orders of
magnitude over the course of the observations, often on timescales as short
as a few hours (e.g. observations A and H). This parameter is
roughly correlated with the iron line equivalent width (Figure
\ref{gxline}a).  At lower $n_{\rm H}$ the EW was also generally lower, but
the results from 11-21 July 1996 (G,H) populate a distinctly different area
of the graph than the majority of spectra, with EW$\la 300$ but $n_{\rm H}$
well above $10^{23}\ {\rm cm}^{-2}$. Also dramatically different was the
point for observation Y2, with $n_{\rm H}\approx 10^{23}\ {\rm cm}^{-2}$ but
EW$\approx1.8$~keV.  Trajectories traced out in $n_{\rm H}-$EW space as the
spectra evolve with time tend to move diagonally, following the mean slope
of the correlation, but there were also significant excursions at various
other angles.  The line energy itself appears to be roughly correlated with
$T_{\rm e}$ (Figure \ref{gxline}b). This does not seem to be a consequence
of a correlation with flux instead, since spectra with widely different
values of flux frequently appear to fall on similar regions of the graph.

\section{Discussion}
\label{c3discussion}

The archival data accumulated through {\it RXTE} observations of GX~1+4
cover a wide range of source conditions for this enigmatic pulsar.
Particularly clear is the degree of variability of the source on all
timescales, from seconds to years.

Analysis of the previously unpublished observations have provided some
important new results. During February 1996 the source was close to the
brightest level seen by {\it RXTE} between 1996 and mid-1997, and the $\approx
1$~d observation at that time featured a dramatic rise in scattering column
density $n_{\rm H}$ over just a few hours.  This event was very similar to
that observed during July 1996 \cite[observation H;][]{gal00:spec}.
Such events are clearly quite common and this strongly suggests $n_{\rm H}$
variations are a significant source of observed countrate modulation on
hourly (and longer) timescales. The modest collecting area of {\it RXTE}
does not provide a high enough countrate to resolve spectral variations on
smaller timescales, but $n_{\rm H}$ variations may have a significant
influence there too.

The iron line energy and the relationship between equivalent width and
$n_{\rm H}$ were roughly consistent with the spherical distribution of matter
inferred by \cite{kot99}. However the minimum timescale of $n_{\rm H}$
variations $\approx 2$~h is much too rapid to be attributable to the
negative feedback effect which those authors suggest regulates mass transfer
to the neutron star in the long term.  The variation may instead be an
indication of significant inhomogeneities in the stellar wind from the
companion on spatial scales of
\begin{equation}
  \delta s \sim 7\times10^9
	\left(\frac{v_{\rm w}}{10\ {\rm km\,s}^{-1}}\right)\ {\rm cm}
\end{equation}
along the line of sight to the neutron star. Infrared observations suggest
that $v_{\rm w}$ may be as much as 250~${\rm km\,s}^{-1}$ \cite[]{chak98:ir},
but since the line of sight may intercept the system at varying distances
form the companion (depending upon the orbital phase) the local wind speed
may also vary due to deceleration.  The rough correlation between $E_{\rm
Fe}$ and $T_{\rm e}$ suggests that there is also a contribution to the Fe
line emission from the accretion column. The relationship between $\tau$ and
$T_{\rm e}$ shows that when the column plasma is cooler it also tends to be
more optically thick; in which case the line emission from the column plasma
will suffer greater self-absorption.  Line emission will thus be dominated
by the cool circumstellar matter, and the resulting line energy is
indicative of very low ionisation levels. At lower source luminosities the
column plasma becomes hotter and slightly more transparent, which makes the
contribution by the more highly ionised column plasma more significant and
increases the fit energy of the gaussian line component. This suggests that
at low luminosities pulse phase dependence of the iron line spectral
parameters may be observed.
Certainly at high luminosities there are suggestions of phase dependence of
e.g. the gaussian component normalisation (Galloway et al. 2000, in
preparation).

\begin{figure}
\plotone{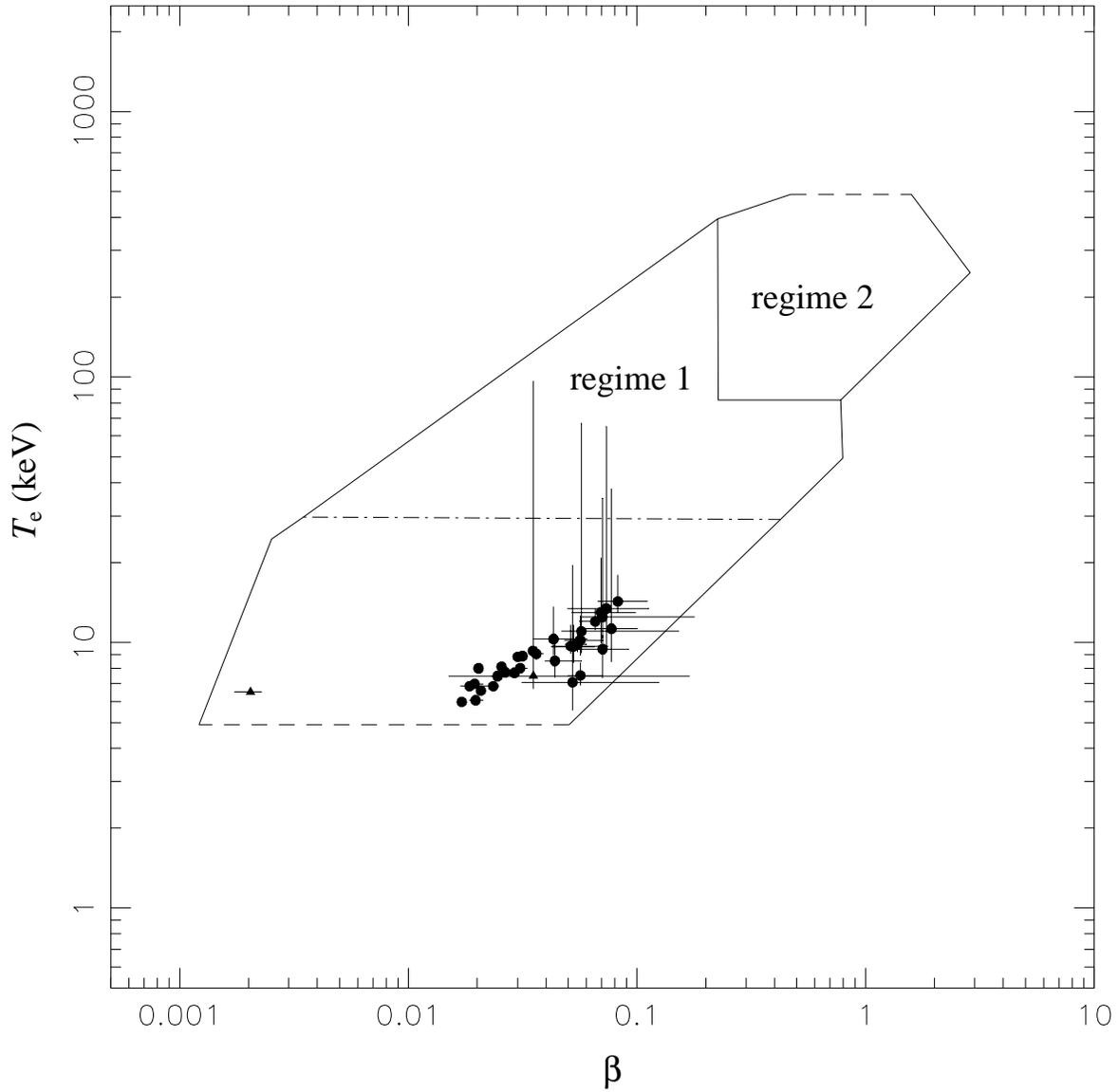}
\caption[gxbeta.eps]{ Scattering plasma temperature $T_{\rm e}$ versus
Titarchuk
$\beta$-parameter appropriate for disc geometries (equation \ref{betaeq}).
The enclosed regions are those plotted in Figure 7 of \cite{ht95}, and
delineate the range of parameters over which the analytical model of
\cite{tit94} is consistent with equivalent Monte Carlo calculations.
Errorbars show 90 per cent confidence limits.
\label{gxbeta} }
\end{figure}

The inclusion of all the archival {\it RXTE} data in the spectral analysis
has confirmed the general applicability of the Comptonization model of
\cite{tit94} to GX~1+4 spectra.
Additionally, model fitting for spectra taken during high-$L_{\rm X}$ intervals
excludes each of the alternative models tested to a high level of statistical
significance.
To assess the reliability of the Comptonization model component
over the range of parameters obtained for GX~1+4, I calculate the
$\beta$-parameter appropriate for disc geometries:
\begin{equation}
  \beta=\frac{\pi^2}{12(\tau+2/3)^2}(1-e^{-1.35\tau})+
	0.45e^{-3.7\tau}\ln\left(\frac{10}{3\tau}\right)
  \label{betaeq}
\end{equation}
\cite[]{ht95}. For the range of $\tau$ and $T_{\rm e}$ appropriate for GX~1+4,
the analytic model has been shown to yield results consistent with
Monte-Carlo simulations \cite[Figure \ref{gxbeta};][]{ht95} and thus should
provide reasonable results. The model simulates Comptonization in an
unmagnetised plasma, and since the available evidence points towards a
strong magnetic field in GX~1+4 (although this awaits confirmation by more
direct measurements such as a cyclotron resonance line) the model fit
parameters may not be an accurate measure of the source conditions. It is
likely that the principal effect of the magnetic field will be to make the
spectral parameters dependent on the emission angle. Hence the model fit
parameters obtained from the mean spectra are expected to be an approximate
measure of the actual physical conditions in the source (L.  Titarchuk 1998,
private communication). 

The source spectrum presumably originates from the polar
cap itself, or perhaps a slab-shaped post-shock region of the column.
Assuming that the majority of the X-ray emission originates from a blackbody
at most the size of the neutron star ($R_* \approx 10$~km) the blackbody
temperature is $T_0 \gtrsim 0.5$~keV. The model fit values for the input
spectrum temperature $T_0 \approx 1.3$~keV was consistent with this
calculation. The fit values for $T_0$ were remarkably consistent for the
majority of spectra. The observation of a much lower value during
observation A was the exception but the errors were quite large. (It is
possible the latter measurements were related to a `transitional' spectrum
similar to that from observation Y2). The polar cap temperature is
presumably maintained by the supply of energy from residual thermal and
kinetic motion of the accreting plasma in competition with cooling by
radiation and conduction through the neutron star crust. That the
temperature varies little over the range of source luminosities points to
extremely efficient cooling mechanisms.

\begin{figure}
\plotone{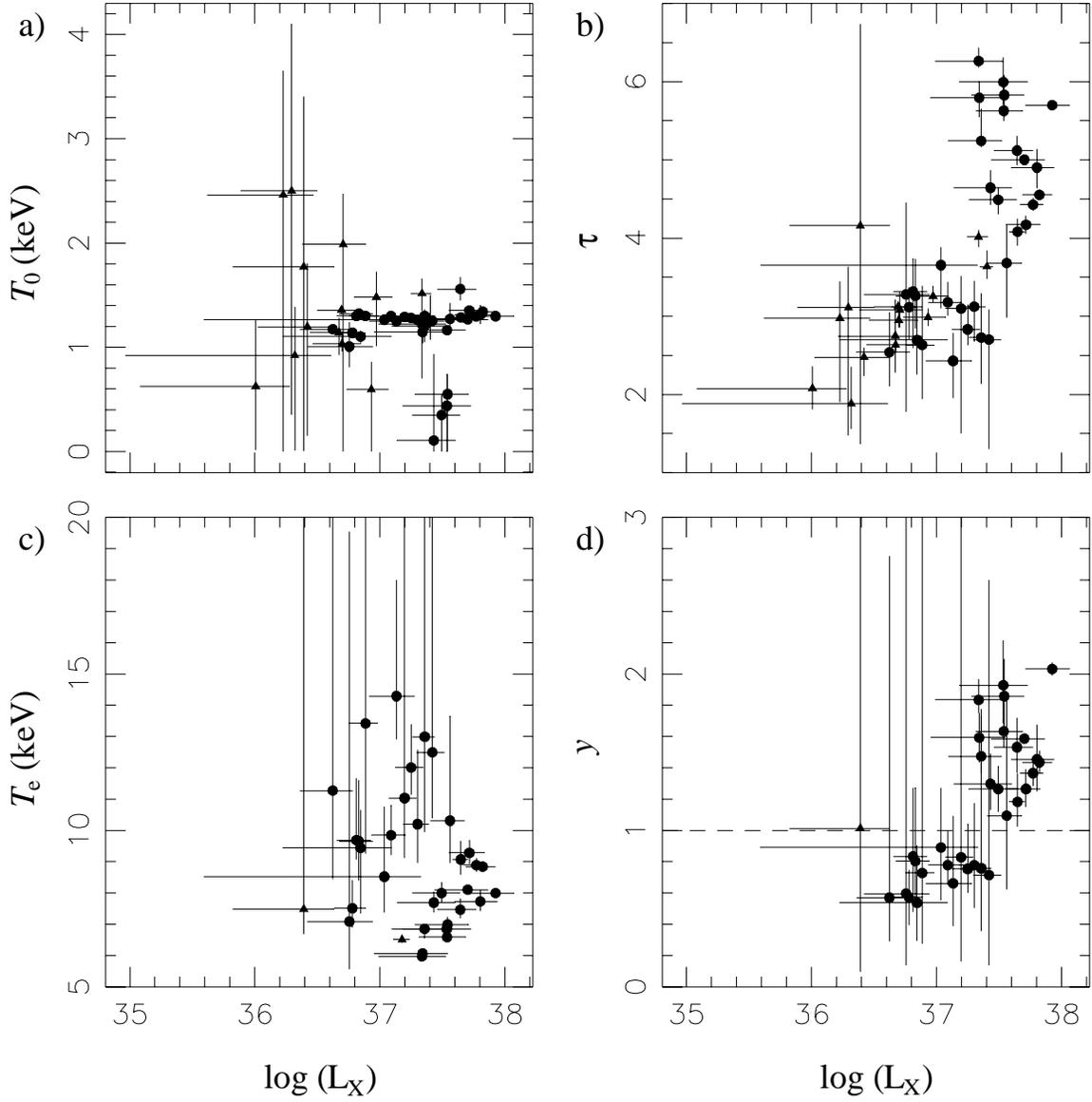}
\caption[comp-param]{ Fit values for the Comptonization continuum
component (`{\tt
compTT}' in {\sc xspec}) parameters for spectral fits to {\it RXTE}
observations of GX~1+4 versus unabsorbed 2--60~keV flux (assuming a source
distance of 10~kpc). a) Source spectrum temperature $T_0$; b) optical depth
for scattering $\tau$; c) plasma temperature $T_{\rm e}$; and d)
Compton $y$-parameter (calculated from the fitted $\tau$ and $T_{\rm e}$
values).  Errorbars show 90 per cent confidence limits. Other details are as
for Figure \ref{gxchi}.
\label{gxcomp}}
\end{figure}

Rough estimates of the accretion column density can be made based
on the mass transfer rate derived from the luminosity, and assuming a simple
column geometry. The accretion luminosity
$L_{\rm acc} \approx GM_* \dot{M}/R_*$ and thus
\begin{eqnarray}
  \dot{M} & \approx & \frac{R_*}{GM_*} L_{\rm acc} \nonumber \\
          & =       & 5.0 \times 10^{16}\,L_{37}\ {\rm g\,s}^{-1}
\end{eqnarray}
where $L_{\rm acc}=L_{37}\times10^{37}\ {\rm erg\,s}^{-1}$.  Assuming that the
accretion column is homogeneous and cylindrical with radius $R_{\rm C}=fR_*$,
and the column plasma is moving with constant velocity $v$ (roughly equal to
the free fall velocity $\approx 0.5c$), the estimated optical depth for
Thompson scattering is
\begin{eqnarray}
  \tau & = & R_{\rm C} \sigma_{\rm T} N_{\rm e} \nonumber \\
       & = & \frac{2 \dot{M} \sigma_{\rm T}}{\pi m_{\rm p} fR_* c} \\
       & = & 2.2\,L_{37} \left( \frac{f}{0.04} \right)^{-1}
                        \left( \frac{v}{c} \right)^{-1} \nonumber
\end{eqnarray}
where $m_{\rm p}$ is the proton mass.  In general $f$ is subject to
considerable uncertainties \cite[e.g.][]{fkr92}, and will depend on the
inner disc radius since that defines which magnetic field lines form the
edge of the accretion column at the neutron star surface. It is usually 
thought that $f=0.01$--$0.1$, with smaller values appropriate for higher
magnetic field systems (where the disc is disrupted further out from the
neutron star on average).  Regardless of the actual values, for the {\it
RXTE} archival data with $L_{37} \in (0.1,10)$ {\em constant} $f$
implies that $\tau$ will also vary by around 2 orders of magnitude. The
observed range of $\tau$ was dramatically different (Figure \ref{gxcomp}b)
and thus it is clear that there are significant changes in accretion column
structure as $L_{\rm X}$ varies.

The small range in $\tau$ (despite the two order-of-magnitude variation
in $L_{\rm X}$) points to a partial regulation of the column density,
possible through the following mechanism. As the accretion rate (and hence
the luminosity) increases, the ram pressure from the gas in the accretion
disc causes the accreting matter to thread magnetic field lines with
footpoints at progressively lower (magnetic) latitudes on the star.  The
extent of the accretion column ($f$) will increase, offsetting the density
increase due to higher $\dot{M}$. As the luminosity increases, the effect of
radiation pressure will also increase.  As pointed out by \cite{lang82}, the
effect of radiation pressure depends strongly on the isotropy of the
radiation, since photons propagating at right angles to the magnetic field
cannot affect the momentum of the plasma flow which is restricted to the
local field direction.  Nevertheless, an increased radiation pressure would
reduce the flow velocity, increasing the density of the flow and hence the
optical depth for scattering. From these arguments it seems much more likely
that $f$ was changing significantly as the accretion rate varied, and the
contribution of radiation pressure was less important. This implies that over
the observed flux range for the {\it RXTE} data $f$ must vary by a factor of
$\approx 30$--50 in order to give the observed variation in $\tau$.

Both $\tau$ and $T_{\rm e}$ appear to be roughly correlated with $L_{\rm
X}$, the latter inversely. This correlation may partially arise from
grouping of the fit values around significantly different means above and
below $L_{\rm X}\approx2\times10^{37}\ {\rm erg\,s}^{-1}$, since the
correlation does not seem to be present in the high- or low-flux
observations fitted separately. As discussed previously, this suggests two
distinct spectral states for the source, although the physical differences
between the two states are not clear.  The combined variation of $\tau$ and
$T_{\rm e}$ result in the calculated Compton $y$-values forming two groups
which are separated by the line $y=1$ (although the uncertainty was generally
rather large). It should be noted that there was significant overlap in flux
between the two groups, even neglecting the uncertainties which were also
present in the flux estimations.  This provides further evidence that some
additional factor may give rise to the observed variation. It is compelling
to suggest that the switch is due to a change from a situation when
Comptonization is relatively unimportant to one where it is the dominant
effect in spectral formation, but such distinctions are really only
appropriate for the asymptotic cases $y\ll 1$ and $y\gg 1$ respectively.
Additionally there is no way to estimate the systematic errors on any of the
parameters involved, and so the fact that the $y$-values cross the $y=1$
line as $L_{\rm X}$ increases was perhaps merely a coincidence.
Nevertheless, the jump may indeed be due to a switch to a significantly
different accretion column structure. One possibility is that the accretion
column becomes much less homogeneous. As the accretion rate increases,
threading of the accreting plasma will become less efficient. Above $L_{\rm
X}\approx2\times10^{37}\ {\rm erg\,s}^{-1}$ large blobs of relatively dense
plasma may survive the accretion process without becoming threaded by
magnetic field lines and thus dispersed, resulting in a
significantly inhomogeneous column which may contribute to the greater
scatter in $\tau$ at high $L_{\rm X}$.  This inhomogeneity may introduce
quasiperiodic oscillations similar to those observed in Cen X-3
\cite[]{jern00}. A search for such oscillations in GX~1+4 at varying mean
flux levels would be a powerful test of this hypothesis.

The $T_{\rm e}$ parameter can be interpreted as a measure of the temperature
of the accretion column plasma. However, the kinetic energy of the plasma
from the infall velocity may be many orders of magnitude larger than the
thermal energy. Photons undergoing Compton scattering in a plasma moving
relative to the source will be scattered preferentially in the direction of
motion, and the influence of the plasma temperature in its rest frame for
the final photon distribution may be small. Thus the fitted value of $T_{\rm
e}$ may owe more to the bulk plasma velocity than its temperature. With
the significant uncertainties in the fit values it is difficult to make any
qualitative statements about the variation with $L_{\rm X}$, but a net
decrease in $T_{\rm e}$ with increasing $L_{\rm X}$ was the most robust
result, possibly related to similar grouping of values as was observed with
the $\tau$ values.  The model normalisation parameter $A_{\rm C}$ is
somewhat more difficult to relate to a physically measurable quantity, since
both the $T_{\rm e}$ and $\tau$ parameters can also affect the total flux
from the model component. Clearly in the absence of spectral variation it
will generally be proportional to the source flux.  

\acknowledgments

This research has made use of data obtained through the
High Energy Astrophysics Science Archive Research Center Online Service,
provided by the NASA/Goddard Space Flight Center, and also the BATSE
Pulsar Group WWW page at \url{ http://www.batse.msfc.nasa.gov/batse/pulsar}.
The {\it RXTE} Guest Observer Facility provided timely and vital help and
information throughout.






\clearpage







\clearpage

\begin{deluxetable}{lcccl}
\tabletypesize{\scriptsize}
\tablecaption{ {\it RXTE} observations of GX 1+4. Start and end times and
on-source durations are calculated from Standard-2 spectra taking into account
screening.  Times are terrestrial time (TT).
    \label{tab-gxobs} }
\tablewidth{0pt}
\tablehead{
  \colhead{ID} & \colhead{Start} & \colhead{End} & \colhead{On-source (sec)}
     & \colhead{Ref.} }
\startdata
   A & 12/02/1996 14:11:11 & 13/02/1996 13:35:11 & 46816 & \\ 
   B & 17/02/1996 03:12:15 & 17/02/1996 08:12:02 & 10400 & [1] \\ 
   C & 23/04/1996 19:10:23 & 23/04/1996 22:40:15 & 7680 & [1] \\ 
   D & 21/05/1996 15:45:03 & 21/05/1996 18:08:15 & 5568 & [1] \\ 
   E & 08/06/1996 04:49:14 & 08/06/1996 11:58:34 & 10112 & [1] \\ 
   F & 28/06/1996 06:41:03 & 28/06/1996 23:35:35 & 5392 & [1] \\ 
   G & 11/07/1996 18:23:27 & 11/07/1996 22:08:40 & 3552 & [1] \\ 
   H & 19/07/1996 16:47:16 & 21/07/1996 02:39:02 & 52832 & [2,3] \\ 
   I & 04/09/1996 22:33:35 & 05/09/1996 03:05:00 & 9232 & [1] \\ 
   J & 25/09/1996 11:12:15 & 25/09/1996 15:19:04 & 6672 & [1] \\ 
   K & 08/10/1996 09:43:32 & 08/10/1996 10;38:44 & 3280 & [1] \\ 
   L & 16/10/1996 18:32:38 & 16/10/1996 23:07:11 & 6016 & [1] \\ 
   M & 24/10/1996 17:58:39 & 24/10/1996 19:51:27 & 4528 & [1] \\ 
   N & 01/11/1996 03:35:27 & 01/11/1996 05:34:02 & 4720 & [1] \\ 
   O & 07/11/1996 18:26:00 & 07/11/1996 20:23:59 & 4592 & [1] \\ 
   P & 10/11/1996 12:16:16 & 10/11/1996 14:13:02 & 4688 & [1] \\ 
   Q & 16/01/1997 00:27:11 & 16/01/1997 02:24:03 & 4592 & [1] \\ 
   R & 16/01/1997 03:16:35 & 16/01/1997 08:02:04 & 9760 & \\ 
   S & 21/01/1997 19:13:51 & 21/01/1997 21:32:31 & 5152 & [1] \\ 
   T & 26/01/1997 22:57:03 & 27/01/1997 00:54:15 & 4672 & [1] \\ 
   U & 02/02/1997 14:35:11 & 02/02/1997 16:50:02 & 4512 & [1] \\ 
   V & 26/02/1997 00:41:15 & 27/02/1997 18:48:31 & 2816 & \\ 
   W & 20/03/1997 16:39:00 & 25/03/1997 18:56:02 & 4768 & \\ 
   X & 18/04/1997 12:34:28 & 22/04/1997 23:45:02 & 4592 & \\ 
   Y & 16/05/1997 03:35:27 & 20/05/1997 07:04:36 & 4080 & \\ 
\enddata
\tablerefs{
[1] \cite{cui97}; [2] \cite{gal00:spec}; [3] \cite{gil00} }
\end{deluxetable}



\end{document}